\documentclass{andromedaone}       

\journal{BSM}
\vol{2021}
\jyear{Egypt}
\pages{Zewail City} 

\def\be{\begin{equation}}
\def\ee{\end{equation}}
\def\bea{\begin{eqnarray}}
\def\eea{\end{eqnarray}}

\def\simg{{\ \lower-1.2pt\vbox{\hbox{\rlap{$>$}\lower6pt\vbox{\hbox{$\sim$}}}}\ }}
\def\siml{{\ \lower-1.2pt\vbox{\hbox{\rlap{$<$}\lower6pt\vbox{\hbox{$\sim$}}}}\ }}
\newcommand{\rmi}[1]{{\mbox{\scriptsize #1}}}

\usepackage{slashed}
\usepackage{bm}

\begin{document}

\title{Revisiting freeze-in dark matter from  renormalizable operators}

\author{Simone Biondini}
\address{Department of Physics, University of Basel,
\\
 Klingelbergstr. 82, CH-4056 Basel, Switzerland}

\begin{abstract}
In this conference paper we summarize the findings of a recent study \cite{Biondini:2020ric}, where the impact of the ultra-relativistic regime on the production of a feebly interacting dark matter particle is considered. As its population accumulates over the thermal history, we inspected thoroughly the temperature window $T \gg M$, which has been previously neglected in the context of dark matter models with renormalizable operators. At high temperatures, and for the model considered in our work, the production rate of the feebly interacting  particle is driven by multiple soft scatterings, as well as $2 \to 2$ processes, that can give a large contribution to the dark matter energy density.
\end{abstract}

\maketitle

\vspace{0.5 cm}
\section*{Introduction}
 It is well possible, despite not the only viable option,  that dark matter (DM) comes in the form of a particle, either elementary or composite.  This possibility opens up a deep and intriguing connection between particle physics and cosmology.  Many theories beyond the Standard Model feature viable candidates for dark matter (see e.g.~\cite{Bertone:2016nfn}).  However, for any model one may want to consider, it has to reproduce a rather accurate measurement: the present-day energy density for DM, $ (\Omega_{\hbox{\scriptsize DM}}h^2)_{\hbox{\scriptsize obs.}}=0.1200 \pm 0.0012$~\cite{Aghanim:2018eyx},  where $h$ is the reduced Hubble constant.  The connection between the model Lagrangian in terms of fields and parameters  (namely masses and couplings),  and the energy density is obtained by means of a \emph{production mechanism} in the early universe.  In the case of the well and widely studied freeze-out mechanism, DM particles share sizeable interactions with the thermal bath, and this typically implies many viable detection strategies.   
A key assumption of the freeze out mechanism is the thermal equilibrium (chemical and kinetic) between DM and the plasma constituents.  Weakly Interacting Massive Particles (WIMPs) belong to this class of DM candidates and the experimental searches have been cornering the parameter space of many models, which are now put under tension \cite{Arcadi:2017kky}. Perhaps this very fact has triggered a renewed interest for another class of DM candidates: feebly interacting massive partcles (FIMPs).  At variance with WIMPs, their interactions with the surrounding plasma is so weak that FIMPs never reached thermal equilibrium.  The population of a FIMP,  if starting with a vanishing one after reheating,  is entirely generated during the thermal history by $2 \to 1 $,   $1 \to 2 $ and $2 \to 2 $ processes, where the FIMP appears \emph{only} in the final state.  The  production mechanism for feebly interacting particles in the early universe is called \textit{freeze-in} \cite{McDonald:2001vt,Hall:2009bx} (see\cite{Baer:2014eja,Bernal:2017kxu} for reviews on the topic and possible detection strategies). 

Let us stress that the relevant temperature range for  freeze-in is complementary with respect to that for freeze-out. This very fact holds for models with renormalizable interactions, where the dark matter production is dominated by $T \sim M$ \cite{McDonald:2001vt,Hall:2009bx} (also dubbed as \emph{infra-red} freeze-in), as well as when non-renormalizable interactions are involved. In the latter case, usually referred to as \emph{ultra-violet} freeze-in \cite{Hall:2009bx,Yaguna:2011ei,Krauss:2013wfa,Elahi:2014fsa}, the production mechanism is sensitive to much higher temperatures, such as the reheating temperature which is set by reheating/end of inflation dynamics. 

In this conference paper, we shall briefly review the findings of a recent study \cite{Biondini:2020ric}, where it is shown how the high-temperature contribution to the FIMP production can be very important (namely $T \gg M$), even in the case one deals solely with renormalizable operators.  The physical situation is very similar to  the
production/equilibration rate of Majorana neutrinos
in type-I seesaw leptogenesis \cite{Anisimov:2010gy,Besak:2012qm,Ghisoiu:2014mha,Ghiglieri:2016xye}, and we  build up on the developments carried out in that research field.  In particular we shall highlight the contribution to the dark matter production rate from multiple soft scatterings, that enhance the $1 \to 2$ decay process and make effective $1\leftrightarrow 2$ processes
possible,  oftentimes called the Landau--Pomeranchuk--Migdal (LPM) \cite{Landau:1953gr,Landau:1953um,Migdal:1956tc}.  We shall consider as well the contribution to $2 \to 2$ processes in the ultra-relativistic regime, namely when the thermal scale $\pi T$ is larger than any other mass scale in the model (in-vacuum and thermal masses). Thermal masses, which are of order $gT$, play a role in both sets of processes, where $g$ here labels the parametrically more important couplings of
the equilibrated degrees of freedom.

In order to illustrate these various effects,  we consider a concrete model with a Majorana dark matter fermion accompanied by a heavier scalar particle, the latter sharing interactions with the Standard Model sector.  More precisely, we consider a simplified model often discussed in the literature \cite{Hall:2009bx,Garny:2017rxs,Belanger:2018sti,Garny:2018icg,Garny:2018ali}, and it belongs to freeze-in models with interesting signatures at colliders.  More precisely,  the dark matter interacts with a SM quark via a colored scalar mediator\footnote{Other realizations are of course possible, where the dark matter particle is a real scalar
or a vector boson \cite{Hisano:2011cs,DiFranzo:2013vra,An:2013xka,Garny:2015wea,Arina:2020udz}.}, and the Lagrangian density reads \cite{Garny:2015wea} 
\\
\\
\begin{eqnarray}
 \mathcal{L} & = & 
 \mathcal{L}^{ }_{\hbox{\tiny SM}} + 
 \frac{1}{2} \, \bar{\chi} \left(  i \slashed{\partial} - M \right)  \chi 
 + (D^{ }_\mu \eta)^\dagger D^\mu \eta 
 - M_\eta^2\, \eta^\dagger \eta 
 - \lambda^{ }_2 (\eta^\dagger \eta)^2 
 \nonumber 
 \\
 & - & \lambda^{ }_3\, \eta^\dagger \eta\, \phi^\dagger \phi 
 - y\,  \eta^\dagger \bar{\chi} a_R q 
 - y^* \bar{q} a_L \chi\, \eta
 \;,
 \label{Lag_RT}
\end{eqnarray}
where $\phi$ is the SM Higgs doublet, $M_\eta$ the mass of the mediator and $M$ the mass of the DM particle, with $M_\eta> M$ so to ensure the fermion being the lightest, and stable, state of the dark sector. The mass splitting between the coloured scalar and the DM fermion is $\Delta M =M_\eta-M >0$. The Yukawa coupling between $\eta$ and $\chi$ is denoted by $y$, whereas $\lambda_2$ and $\lambda_3$  are the self-coupling of the coloured scalar and its portal coupling to the Higgs respectively. The coupling $\lambda_1$ is left for the SM Higgs self interaction and $a_R$ ($a_L$) is the right-handed (left-handed) projector. In this work, we set $\lambda_2=0$ in order to reduce the number  of free parameters of the simplified model.
The heavier mediator is typically responsible for additional dark matter production at much later stages in the thermal history via the super-WIMP mechanism~\cite{Feng:2003xh,Feng:2003uy}. Here, the relic abundance of the mediator, as determined by pair annihilations and thermal freeze-out, is key to the extraction of the dark matter energy density.  Due to QCD interactions experience by the coloured scalar,  we shall include bound-state effects on the late-time annihilations of the colored mediator.

The plan of the paper is as follows.  In section 1 we introduce the particle production rate from a finite-temperature field theoretical point of view, and make contact with the standard Boltzmann equation.  We will consider the Born rate with vanishing and non-vanishing thermal masses as a reference single out the high-temperature effects. These are briefly reviewed in section 2: effective $ 1 + n \to 2 + n $ processes as well as $2 \to 2$ processes in the ultra relativistic regime.   In section 3 we show their impact on the relic energy density of dark matter, after we map out the parameters region where the freeze-in produced component is largely dominant with respect to the Super-WIMP contribution. This assessment is rather important and bound-state effects for the annihilating colored scalars are taken into account. Finally, we offer some concluding remarks in section 4.

\section*{1. Particle production rate and Boltzmann equation} 
\label{sec:part_prod}
The key ingredient in our analysis is the particle production rate, here for the DM particle $\chi$. We consider a quite general approach that allows to obtain the production rate of a weakly coupled particle with an equilibrated bath. The latter has internal couplings denoted with $g$. One can prove that, at leading order in $y$ and at \emph{all} orders in $g$, the rate of change of the single-particle phase space distribution $f_\chi(t, \bm{k})$ reads \cite{Bodeker:2015exa}
\begin{equation}
\label{diff_rate_eq}
\left( \frac{\partial}{\partial t} - H k_i \frac{\partial}{\partial k_i} \right) f_\chi (t,\bm{k})  = \Gamma(k)[n_\mathrm{F}(k^0)-f_\chi (t,\bm{k})],
\end{equation}
where $k^0=\sqrt{k^2+M^2}$, $\mathcal{K}$ is the fermion four-momentum in Minkowski metric and $n_\mathrm{F}$ is the Fermi--Dirac 
distribution. The  production rate $\Gamma(k)$ can be expressed in terms of a two-point correlation function at finite temperature, in our case the self-energy of the Majorana fermion
 \begin{equation}
  \begin{minipage}[c]{0.055\linewidth} \includegraphics[scale=0.42]{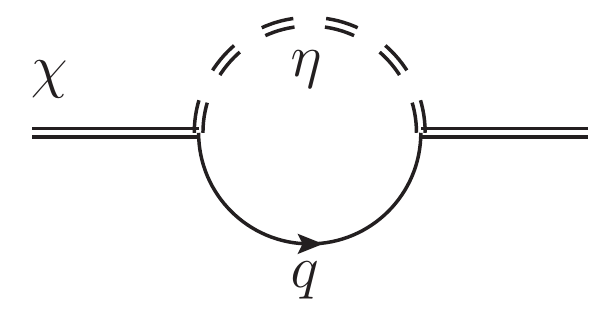} \end{minipage}  \hspace{3.5 cm}
\Gamma(k) =
\frac{|y|^2}{k^0}\mathrm{Im}\Pi_R \, .
\label{defgamma}
\end{equation}
The retarded self-energy comprises equilibrated degrees of freedom (that enjoy a fast dynamics). At leading order in this model, one finds a colored scalar and a SM quark in the one-loop self-energy. As typical of freeze-in production, which features an abundance of the DM much smaller than the corresponding equilibirum one, one has $f_\chi (t,\bm{k}) \ll n_\mathrm{F}(k^0)$. Therefore, we can neglect $f_\chi$ on the right-hand side of the rate equation (\ref{diff_rate_eq}). 
The approach outlined here reproduces a standard Boltzmann equation when applying perturbation theory and working at leading order in the thermal bath coupling $g$. Upon defining the number density of dark matter particles as $n_\mathrm{DM}=2 \int_{\bm{k}} f_\chi(t,k)$, with the factor of $2$ accounting for the two helicity states, we 
rewrite eq.~(\ref{diff_rate_eq}) as follows 
\begin{eqnarray}
\dot{n}_\mathrm{DM} + 3H n_\mathrm{DM} &=& 2 |y|^2 \int_{\bm{k}} \frac{n_\mathrm{F}(k^0)}{k^0}  \textrm{Im} \Pi_R \\
&=& 2 |y|^2N_c (M_\eta^2 -M^2)  \int_{\bm{p}_\eta, \bm{p}_q, \bm{k}}  \frac{(2 \pi)^4 \delta^4(\mathcal{P}_\eta - \mathcal{P}_q - \mathcal{K})}{8 E_\eta E_q \, k^0}  n_\mathrm{B}(E_\eta) \left[ 1-n_\mathrm{F}(E_q)\right] \, ,
\label{number_density_Born}
\end{eqnarray}
where the production rate has been worked out at leading order and simplified according to the sole process that is kinematically allowed in this case: a decay process $\eta \to \chi q$. As one may see on the right-hand side of eq.~(\ref{number_density_Born}), the distribution function of the equilibrated species ($\eta,q$) appear as in a gain term of a Boltzmann equation. The  produced dark matter fermion appears with a unity factor, that conforms with the expectation since $1 + f_\chi \simeq 1$, due to its negligible abundance.  
It is worth noticing that the approach discussed here has the full advantage of a field-theoretical formulation. Then, it offers the possibility to systematically include higher order corrections and thermal effects. 
\subsection*{\textbf{Born rate with vanishing and finite thermal masses}}
In order to set a reference for the high-temperature effects on the production rate and the dark-matter energy density, let us consider the Born rate, namely the leading order contribution to the dark fermion self-energy. The Born rate, with in-vacuum masses, reads 
\begin{eqnarray}
\textrm{Im} \Pi_R^{\hbox{\scriptsize }} \bigg\vert_{\hbox{\scriptsize in-vacuum}}   &=&  
 \frac{N_c T (M_\eta^2 -M^2)}{16 \pi k} \left[ \ln \left( \frac{\sinh(\beta(k^0+p_{\hbox{\tiny max}})/2)}{\sinh(\beta(k^0+p_{\hbox{\tiny min}})/2)}\right) - \ln \left( \frac{\cosh(\beta p_{\hbox{\tiny max}} /2)}{\cosh(\beta p_{\hbox{\tiny min}}/2)}\right)\right] \, , 
 \label{vacuum_born}
 \end{eqnarray}
 with 
  \begin{equation}
     p_{\hbox{\tiny min}}=\frac{M^2_\eta - M^2 }{2(k^0+k)} \, , \quad p_{\hbox{\tiny max}}=\frac{M^2_\eta - M^2 }{2(k^0-k)} \, .
     \label{boundaries_vacuum_born}
 \end{equation}
 and it just amounts at simplifying further the expression that appears in eq.~(\ref{number_density_Born}).
 
 Next, the first non-trivial thermal effects we consider are thermal masses. In a high-temperature environment, the modification to the dispersion relation has to be taken into account and repeated interactions with the plasma constituents generate the so-called asymptotic masses. 
 We write the asymptotic thermal masses for the colored scalar and the SM quark, they read
\begin{eqnarray}
m^2_\eta= \left( \frac{g_3^2C_F+Y_q^2g_1^2}{4} + \frac{\lambda_3}{6}  \right)  T^2 \, , \quad m_q^2= \frac{ T^2}{4}(g_3^2C_F + Y_q^2g_1^2 + |h_q|^2) \, ,
\label{asym_the_mass}
\end{eqnarray}
where we note that the gauge contribution is the same for both the scalar and fermion;  $C_F=(N_c^2-1)/(2N_c)$ is the quadratic Casimir of the fundamental 
representation. The thermal mass for the DM is negligible since it is proportional to $|y|^2 \ll g_3^2, g_1^2, \lambda_3,|h_t|^2$. It is important to stress that, for temperature larger than the SM electroweak symmetry breaking, the only source for the quark mass is indeed the thermal contribution. This will play a role in the estimation of the production rate, since the finite quark mass shrinks the available phase space in the decay process $\eta \to \chi q$, depending on the choice of the parameters and the temperature. Moreover, another crucial difference with the in-vacuum case is that another channel for the dark matter production can be realized, namely $q \to \eta \chi$, again because of the thermal masses. A more comprehensive discussion of the handling of the scalar and quark thermal masses for different temperature regimes can be found in ref.~\cite{Biondini:2020ric}. Here we only give the expression for the Born rate with finite thermal masses for the process $\eta \to \chi + q$ (with ``full'' we mean thermal masses included)
\begin{equation}
\label{fullborneta}
   \textrm{Im} \Pi_R^{\hbox{\scriptsize }} \bigg\vert_{\hbox{\scriptsize full}}   =\frac{N_c}{16\pi k}\int_{p_\mathrm{min}}^{p_\mathrm{max}} d p [\mathcal{M}_\eta^2-M^2-m_q^2-2 k^0 (E_p-p)]  [n_\mathrm{B}(k^0+E_p)+n_\mathrm{F}(E_p)],
\end{equation}
where the total (in-vacuum plus thermal) mass for the scalar is $\mathcal{M}_\eta^2=M_\eta^2+m_\eta^2$, $E_p=\sqrt{p^2+m_q^2}$ and the integration boundaries are
\begin{equation}
    p_\mathrm{min,\,max}=\frac{\mathcal{M}_\eta^2-M^2-m_q^2}{2M^2}\left|k^0\sqrt{1-\frac{4M^2 m_q^2}{(\mathcal{M}_\eta^2-M^2-m_q^2)^2}}\mp k\right|.
    \label{boundaries_thermal_born}
\end{equation}
\begin{figure}
    \centering
    \includegraphics[scale=0.57]{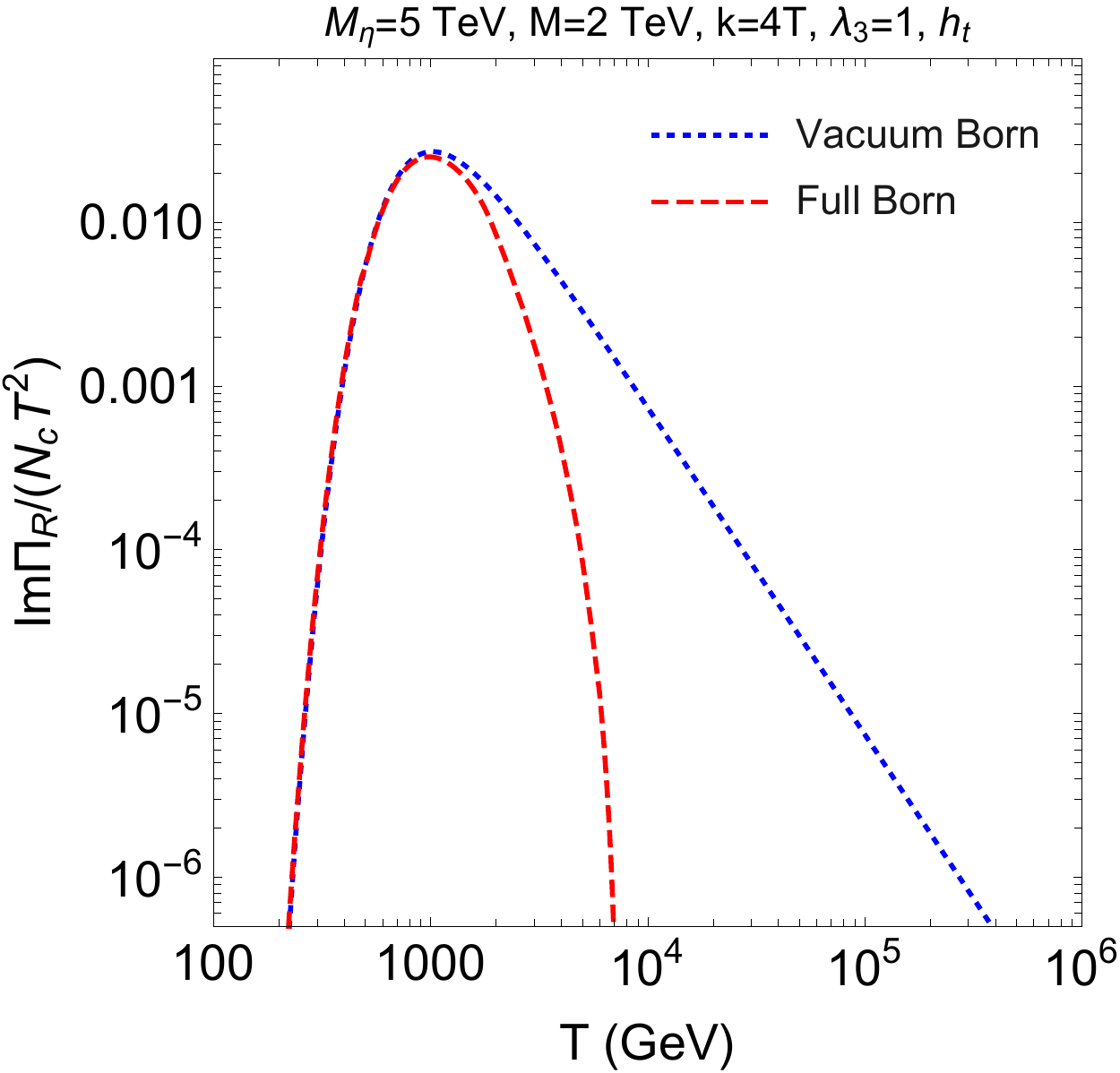}
    \hspace{0.9 cm}
    \includegraphics[scale=0.55]{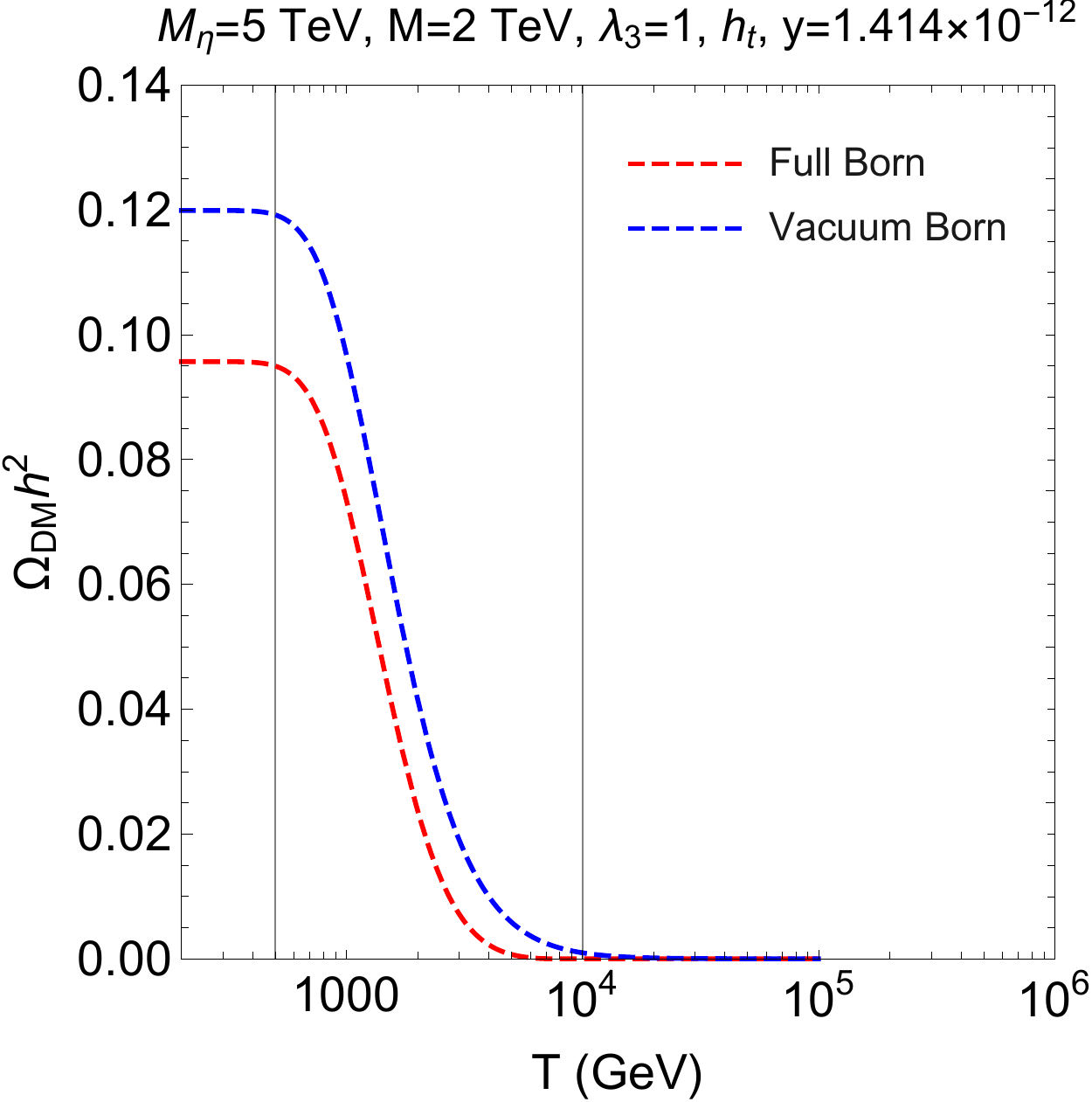}
    \caption{Left panel: the Born rate, divided by $N_cT^2$, computed with in-vacuum masses only (blue-dotted) and with thermal masses included (red-dashed line). Right panel: the dark matter energy density from the Boltzmann equation (\ref{number_density_Born}) with the Born rate with and without thermal masses. The parameters are fixed to reproduce  $(\Omega_{\hbox{\footnotesize DM}}h^2)_{\hbox{\scriptsize obs.}}$ for the in-vacuum Born case. }
    \label{fig:born_vacuum_finite}
\end{figure}

In figure~\ref{fig:born_vacuum_finite} left plot, we show the Born rate with and without thermal masses, red dashed and blue dotted respectively. One may appreciate the effect of finite thermal masses and a smaller rate at high-temperatures due to the finite and large quark thermal mass (for this choice of the parameters). On the right plot, we instead display the evolution of the energy density of the dark fermion as a solution of the Boltzmann equation with the Born rate (\ref{vacuum_born}) and (\ref{fullborneta}). We start with a vanishing initial abundance for $\chi$ and the production of the dark fermion  is driven by the decay process $\eta \to \chi +q$. It is important to notice two aspects. First, the bulk of the dark matter production happens in a temperature window of the order of the decaying particle, here the coloured scalar. The gray vertical lines show the range $M_\eta/10 \leq T \leq 2 M_\eta$ (or $M/4 \leq T \leq 5 M$). For temperature larger than twice the decaying particle, the DM energy density is still close to zero. This shows that, when adopting leading order rates, the bulk of the DM production occurs for temperatures close to the accompanying decaying state mass (or the very same DM mass for small mass splittings). Second, the inclusion of the thermal masses decreases the energy density because the production rate is less effective. 
\section*{2. Ultra-relativistic dynamics} 
\label{sec:ultra_rel}
As we have seen in the previous section, one integrates the Boltzmann equation (\ref{number_density_Born}) from very early times, namely high temperatures, to track the DM production. However, the Born rate, both the in-vacuum (\ref{vacuum_born}) and the one with thermal masses (\ref{fullborneta}), peaks at temperatures of the order of the decaying particle. The production is most effective at temperatures $T \siml M_\eta$. We shall see that there are processes which are instead very effective at high temperatures, and that can change the freeze-in production quite a lot. In particular, at high-temperature, all the particles are essentially seen as  massless with respect to the hard scale $\pi T$ typical of the thermal motion, and hence the angle between the initial state momenta or between
the final state momenta can be small. This defines the collinear kinematics as a distinctive feature of the high-temperature dynamics. 

In the following we briefly discuss two classes of processes responsible for an enhancement of the production rate: 
effective $1\leftrightarrow 2$ processes, induced by multiple soft scatterings, and $2 \leftrightarrow 2$ scatterings, whose correct derivation at leading order need both Hard Thermal Loop (HTL) and LPM resummation. The technicalities are not addressed in this conference paper, and we refer to ref.~\cite{Biondini:2020ric}, and references therein, for a more detailed discussion. 
 
\subsection*{\textbf{LPM resummation and effective $1 + n \leftrightarrow 2 + n$ processes}}
\label{LPM_rate}
Let us start with the definition of hard and soft scales. At high temperatures all external momenta can be taken as hard, $p \sim \pi T$, whereas the thermal masses of the particles define a soft scale of order $gT$. In the regime $T \gg M_\eta,M$, we can treat the vacuum masses $M$
and $M_\eta$ as  soft scales as well. All the particles involved in the reactions are then close to the light cone because $\mathcal{P}^2 \sim (gT)^2$, where $\mathcal{P}$ is a Minkowskian four-momentum.
\begin{figure}[t!]
    \centering
    \includegraphics[scale=0.53]{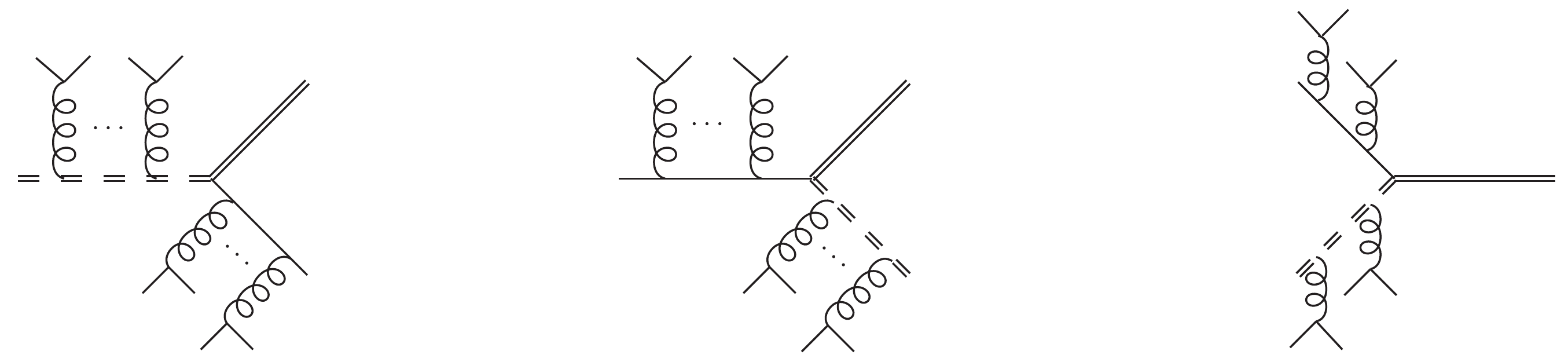}
    \caption{Processes where the QCD gluons induce soft scatterings with plasma constituents (coming from the HTL resummed gluon propagator). Both 
    coloured/charged particles, namely $\eta$ and $q$,  can undergo multiple scatterings. The same processes occur with the U(1)$_Y$ gauge boson.}
    \label{fig:1to2_soft}
\end{figure}

Let us now consider multiple scatterings with plasma constituents, as shown in the three diagrams in figure~\ref{fig:1to2_soft}. Despite these processes seem to be of higher order, due to the many additional vertices, this is not the case. The gauge bosons that are exchanged
with the thermal constituents of the medium are soft, namely $q^0\sim q\sim g T$, and this makes the intermediate virtual $\eta$ bosons and quarks almost on shell, with a lifetime (or \emph{formation time}) of
order $1/g^2T$, which is long and parametrically of the same order of the
soft-gauge-boson-mediated scattering rate (see e.g. \cite{Ghiglieri:2020dpq} for a detailed disucssion). The point is that many of these scatterings occur before the actual emission of the particles in the final states, and their interference has to be taken into account in what is called LPM resummation. In so doing, we speak of \emph{effective} $1\leftrightarrow 2$
processes to describe the $1 + n \leftrightarrow 2 +n$ processes being consistently
accounted for at \emph{leading order}. The gauge bosons propagators (both QCD gluons and $B_\mu$ of U(1)$_Y$) are the HTL ones, in order to properly treat the soft momentum exchange. 

We borrow the notation and computational setting from refs.~\cite{Anisimov:2010gy,Besak:2012qm,Ghisoiu:2014ena,Ghiglieri:2016xye}, where leptogenesis with Majorana neutrinos is studied. The main ingredients to derive the LPM production rate are an effective Hamiltonian $\hat{H}$ that comprises thermal masses for the colored scalar and the quark, as well as the rate that encodes soft gauge scatterings. It is important to stress that the LPM resummation takes care of an arbitrary number of soft scatterings mediated by the QCD gluons as well as the U(1)$_Y$ gauge boson. In this model both possibilities are viable (in figure~\ref{fig:1to2_soft} we only showed soft exchange mediated by gluons), and taken into account in the numerical analysis for the dark matter energy density. Next, the effective Hamiltonian enters the inhomogeneous equations for the functions $g(\bm{y})$ and $\bm{f}(\bm{y})$ that are necessary to compute the LPM effect, that can finally be expressed as 
\begin{eqnarray}
{\rm{Im}}\Pi_{\hbox{\tiny R}}^{\hbox{\tiny LPM}}&=& -\frac{N_c}{8 \pi} \int_{-\infty}^{+\infty} d E_q  \int_{-\infty}^{+\infty} d E_\eta \, \delta(k_0-E_q-E_\eta) [1-n_\mathrm{F}(E_q)+n_\mathrm{B}(E_\eta)] \nonumber 
\\
&& \frac{k_0}{E_\eta} \lim_{\bm{y}\to0} \left\lbrace  \frac{M^2}{k_0^2} {\rm{Im}} [g(\bm{y})] + \frac{1}{E_q^2} {\rm{Im}} [\nabla_\perp \cdot \bm{f}(\bm{y})]\right\rbrace \, . 
\label{lpm_eq}
\end{eqnarray} 
It is worth pointing out that the integration region in eq.~(\ref{lpm_eq}) accounts both for
effective $1\to 2$ processes and for effective $2\to 1$ processes. One may see this by fixing $E_\eta$ 
to $k^0-E_q$ by the $\delta$ function, and then three distinct regions are viable\footnote{In this list
we do not distinguish between particle and antiparticle states. Strictly speaking we should have e.g. $\bar{\eta},q\to \chi$.}
\begin{enumerate}
    \item  $k^0>E_q>0$: this corresponds to the effective $2\to 1$ process $\eta,q\to \chi$,
    \item \label{etadecay} $E_q<0$: this corresponds to the effective $1\to 2$ process $\eta\to q \chi$. eq.~ref{collborn} is the $n=0$ limit (no scatterings) thereof,
    \item $E_q>k^0$: this corresponds to the effective $1\to 2$ process $q\to \eta \chi$.
\end{enumerate}
Without accounting for soft scatterings, at most one of these three scenarios is realized at a 
time, e.g. scenario~\ref{etadecay} for $\mathcal{M}_\eta>m_q+M$. The inclusion of soft
scatterings makes all three options kinematically
allowed simultaneously.

\begin{figure}[t!]
    \centering
    \includegraphics{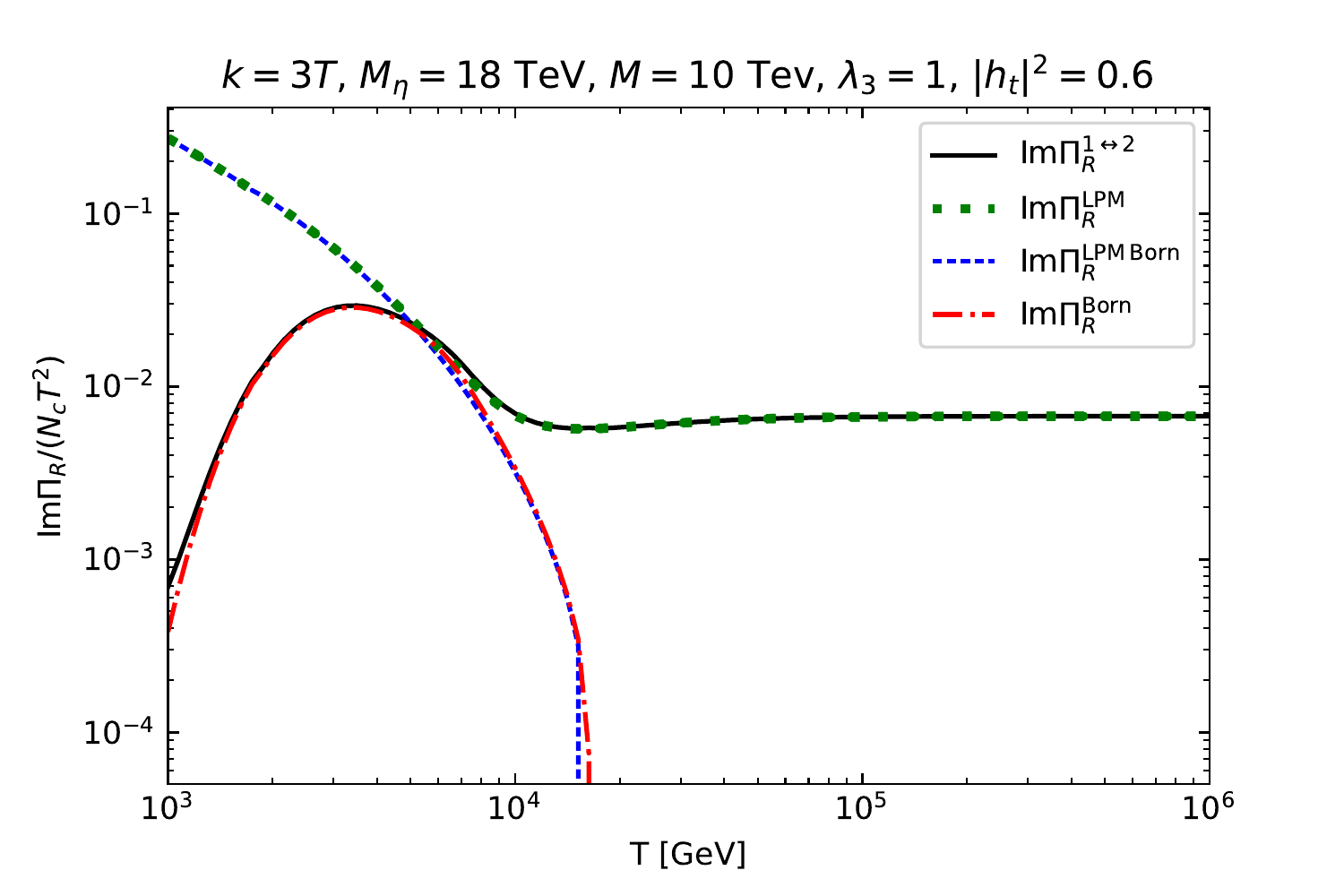}
    \caption{In this figure, the strong coupling is fixed at $g_3^2=0.85$, and the effect of the U(1)$_Y$ gauge coupling is neglected, i.e. $g_1=0$.
    The values
    for the masses listed above the plot are the zero-temperature masses.
    }
    \label{fig_subtr_ht}
\end{figure}

As we show in figure~\ref{fig_subtr_ht}, the effective $1\leftrightarrow 2$ processes give the dominant rate at high temperatures (dotted green line), that drives the dark fermion production at temperatures much larger than the colored scalar mass. The dot-dashed red curve stands for the Born rate with thermal masses included (\ref{fullborneta}), whereas the blue-dotted line represents the collinear limit. The latter is computed from the LMP rate by imposing no soft scatterings. As shown in the plot, there is an enhancement at small temperatures which is however non-physical: the collinear limit fails because of the temperature becoming comparable with the in-vacuum masses. Here the sensitivity to the in-vacuum scales has to be reinstated, and our best effort curve is the black solid line that corresponds to 
\begin{equation}
    \label{sub_pres}
    \mathrm{Im}\Pi_R^{1\leftrightarrow 2}=\mathrm{Im}\Pi_R^\mathrm{LPM}-
    \mathrm{Im}\Pi_R^\mathrm{LPM\;Born}+\mathrm{Im}\Pi_R^\mathrm{Born}.
\end{equation}
which ensures a reasonable $1\leftrightarrow2$ rate at all values of $M/T$ and could be ameliorated by the knowledge of the relativistic NLO dark fermion self-energy (at the moment not available, see refs.~\cite{Laine:2013vpa,Laine:2013lka} for the present state-of-art). 
\subsection*{\textbf{$2 \to 2$ scatterings}} 
\label{2to2_rate}
\begin{figure}[t!]
    \centering
    \includegraphics[scale=0.5]{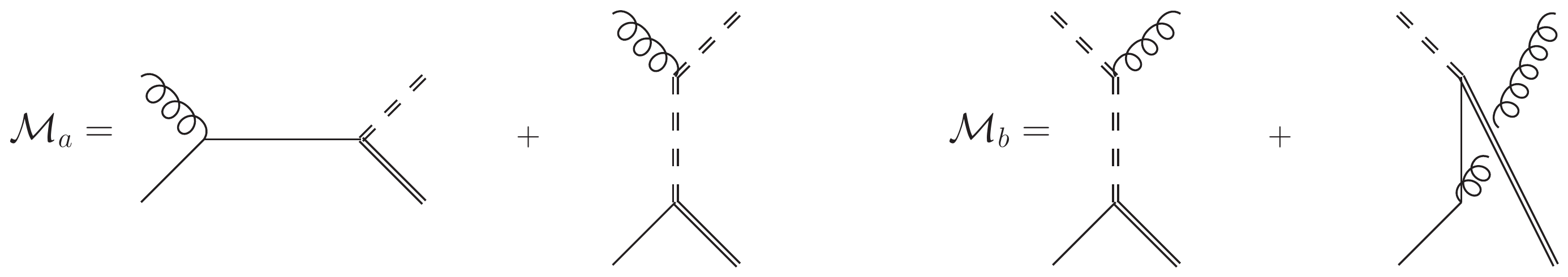}
    \caption{The diagrams contributing to the $2 \to 2$ processes $g \, q \to  \eta \chi$ and $ \eta^\dagger q \to g \chi$ are shown. The same diagrams with a U(1)$_Y$ gauge boson that replaces the QCD gluon (curly line) contribute as well. The complete list of diagrams is given in ref.~\cite{Biondini:2020ric}.}
    \label{fig:2to2_ab}
\end{figure}

In addition to $1 \to 2$ effective processes, there are also $2 \to 2$ scattering processes to be considered. In this model, there are many diagrams, an example of which is given in figure~\ref{fig:2to2_ab}. In the high-temperature regime, they are expected to contribute at the same order of multiple soft scatterings. We notice that $2 \to  2$ scatterings have been already considered in ref.~\cite{Garny:2018ali}, though for $T\siml M$
and possibly with some limitations due to a proper handling of the IR divergences in some of the diagrams.  Indeed there are two distinct momentum regions for the momentum transfer that contribute to the $2 \to 2$ scatterings: a hard region $q \sim \pi T$ and a soft region $q \sim g T$. As for the former, the leading order equivalence with a Boltzmann equation holds and we can use eq.~(\ref{diff_rate_eq}) to obtain 
\begin{eqnarray}
  \label{kin_thy}
  \dot{f}_\chi (k)&=& n_\mathrm{F}(k) \Gamma(k) \bigg\vert_{2\leftrightarrow 2}^\mathrm{hard}  \, + \cdots
  \nonumber 
  \\
  &=&\frac{1}{4k}
  \int \! {\rm d}\Omega^{ }_{2\leftrightarrow2} \sum_{abc} 
  \Bigl\vert\mathcal{M}^{ab}_{c\chi}
  (\bm{p}_1,\bm{p}_2;\bm{k}_1,\bm{k})\Bigr\vert^2 
  f^{ }_a(p^{ }_1)\,f^{ }_b(p^{ }_2)\,[1\pm f^{ }_c(k^{ }_1)]  \, + \cdots
  \;,
\end{eqnarray}
where we have again neglected $f_\chi (k)$ for the dark matter appearing in the final state and the ellipses stand for other production processes, such as the $1\leftrightarrow 2$ channels. The summation runs over all possible $2 \to 2$ processes. As for the soft region,  one has to properly handle the sensitivity to such momentum scale for processes involving a fermionic $t$-channel. This can be achieved by implementing a HTL resummation for the quark propagator. This way, one obtains a finite and physical result.  We implement the subtraction as detailed in  ref.~\cite{Ghiglieri:2016xye}, and the final result reads
\begin{eqnarray}
\textrm{Im}\Pi_R^{2\leftrightarrow 2} & = & 
 \frac{2}{(4\pi)^3k}
 \int_{k}^{\infty} \! {\rm d} q_+ \int_0^{k} \! {\rm d} q_- 
 \Bigl\{ 
  \bigl[ n_\mathrm{F}{}(q_0) + n_\mathrm{B}{}(q_0 - k)  \bigr] 
  N_c\bigl( Y_q^2g_1^2+ C_F g_3^2+ \vert h_q\vert^2\bigr) \,  \Phi_\rmi{$s$2} 
 \Bigr\} 
 \nonumber \\
 & + & 
 \frac{2}{(4\pi)^3k}
 \int_{0}^{k} \! {\rm d} q_+ \int_{-\infty}^{0} \! {\rm d} q_-  
 \Bigl\{ 
  \bigl[1 - n_\mathrm{F}{}(q_0) + n_\mathrm{B}{}(k - q_0) \bigr]   
 N_c\bigl( Y_q^2g_1^2+ C_F g_3^2+ \vert h_q\vert^2\bigr)  \,\Phi_\rmi{$t$2} 
 \nonumber\\
 & & 
  -  \, \Bigl[n_\mathrm{B}{}(k) + \frac12\, \Bigr]
 \,  
 N_c\bigl( Y_q^2g_1^2+ C_F g_3^2+ \vert h_q\vert^2\bigr) \,  \frac{k \pi^2 T^2}{q^2}
 \, \Bigr\} 
 \nonumber \\
 & + & 
 N_c\frac{m_q^2}{16\pi} \, \Bigl[n_\mathrm{B}{}(k) + \frac12\, \Bigr] 
 \, \ln\left(1+ \frac{4k^2}{m_q^2} \right)
 \; + \; 
 \mathcal{O}\Bigl( \frac{m_q^4}{k^3} \Bigr)
 \;. \label{direct_full}
\end{eqnarray}
where the expressions for the $s$- and $t$ channel functions $\Phi_{s2}$ and $\Phi_{t2}$ is given in ref.~\cite{Biondini:2020ric}.  
\begin{figure}[t!]
    \centering
    \includegraphics{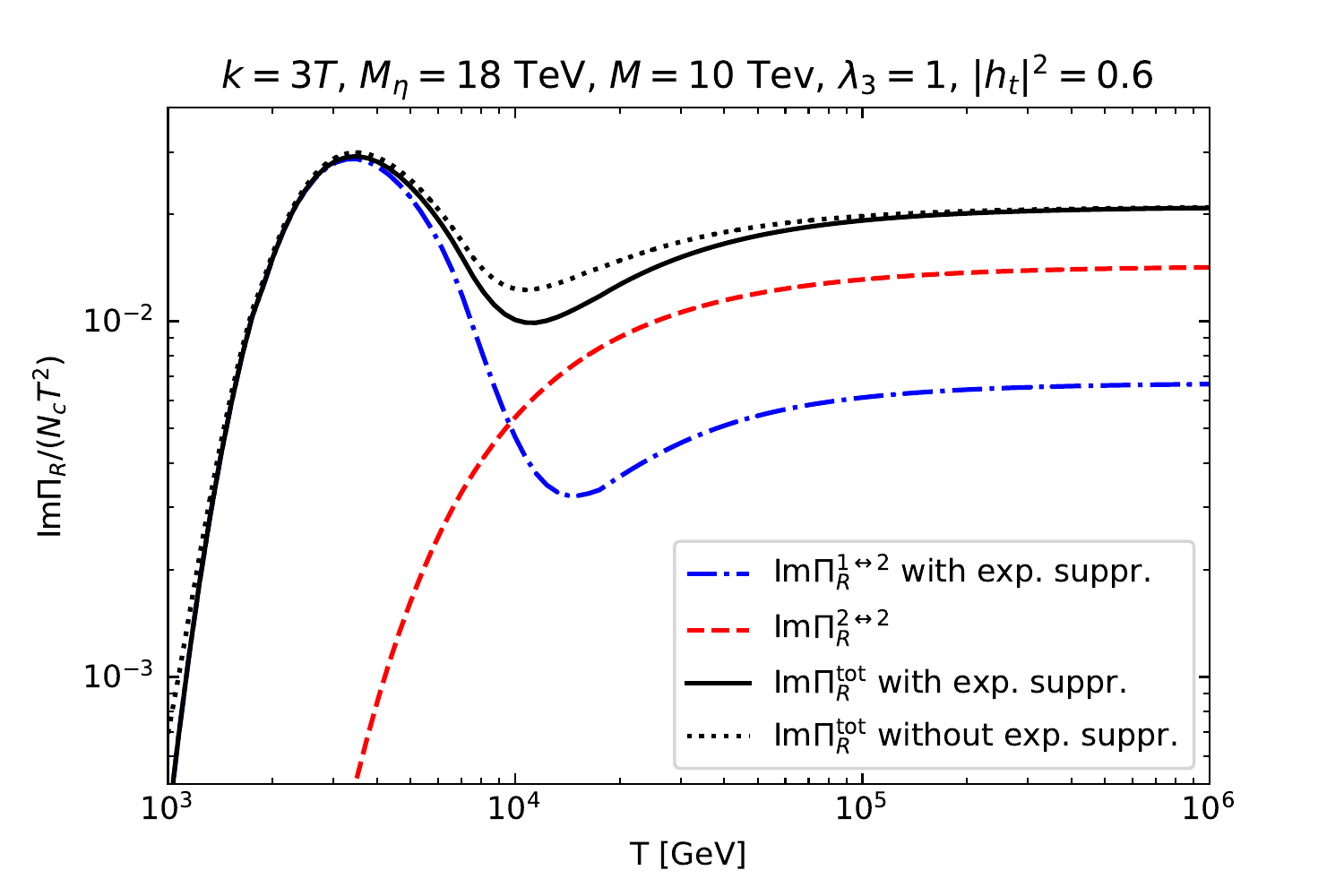}
    \caption{The $1\leftrightarrow2$ rate in eq.~(\ref{sub_pres_2}) and $2\leftrightarrow2$ in eq.~(\ref{direct_full}),
times the susceptibility factor $\kappa(M_\eta)$. The solid black line is their sum as in eq.~(\ref{defpitot}) (the dotted-black curve stands for the $1\leftrightarrow2$ without $\kappa(M_\eta)$).}
    \label{fig_12_22}
\end{figure}
\subsection*{\textbf{Summary of the rates and phenomenological prescription}}
In preparation for the next section, we need to complement the treatment of the high-temperature processes in order to follow the entire production process down to smaller temperatures. The main point is that, while the universe cools down, the dynamics of the production processes is increasingly affected by the in-vacuum masses.  As we assumed $k\sim \pi T \gg M,M_\eta$, the calculations presented in the former section cease to
be valid for $T \siml M,M_\eta$, where the full Born term
of eq.~(\ref{fullborneta}) is the leading-order term and subleading 
corrections are unknown in the relativistic regime. 

In order to switch off such rates progressively when
approaching this region in our analyses, we follow the recipe
inspired by ref.~\cite{Ghiglieri:2016xye}: we shall multiply $(\mathrm{Im}\Pi_R^{\hbox{\tiny LPM}} -\mathrm{Im}\Pi_R^{\hbox{\tiny Born LPM}})$ and $\mathrm{Im}\Pi_R^{2\leftrightarrow2}$ by a factor $\kappa(M_\eta)$, which is obtained
by taking the $\eta$ boson susceptibility normalised by its massless limit. The susceptibility factor reads
\begin{equation}
    \label{defkappa}
    \kappa(M_\eta)=\frac{3}{\pi^2T^3}\int_0^\infty dp\,p^2\,n_\mathrm{B}(E_\eta)
    [1+n_\mathrm{B}(E_\eta)]\,.
\end{equation}
Even though this prescription is based on a phenomenological argument rather than a rigorous implementation, it captures well the sensitivity to the largest in-vacuum mass scale, namely $M_\eta$. In figure~\ref{fig_12_22}, we compare and sum 
our final $1\leftrightarrow2$ results with
our $2\leftrightarrow2$ taking into account
the phenomenological switch-off in eq.~(\ref{defkappa}), and we define the total rate as
\begin{equation}
    \mathrm{Im}\Pi_R^\mathrm{tot}=  \mathrm{Im}\Pi_R^{1\leftrightarrow 2}+
\label{defpitot}
     \mathrm{Im}\Pi_R^{2\leftrightarrow 2} \, ,
\end{equation}
where the susceptibility factor enters the collinear LPM rates as follows
    \begin{equation}
    \label{sub_pres_2}
    \mathrm{Im}\Pi_R^{1\leftrightarrow 2}=(\mathrm{Im}\Pi_R^\mathrm{LPM}-
    \mathrm{Im}\Pi_R^\mathrm{LPM\;Born})\kappa(M_\eta)+\mathrm{Im}\Pi_R^\mathrm{Born} \, .
\end{equation}

\section*{3. Impact on the relic density}
\label{sec:rel_density}
For the model at hand, there are two contributions to the energy density of dark matter \cite{Garny:2018ali,Arcadi:2013aba}: the freeze-in mechanism, that dominates at temperatures $T \simg M_\eta$, and the super-WIMP mechanism~\cite{Feng:2003xh,Feng:2003uy}, that instead takes place much later at $T \ll M_\eta$. In the latter case, the freeze-out of the $\eta$ particles occurs similarly to a WIMP (despite the leading interactions are driven by the strong coupling $g_3$), and dark matter fermions are produced in the subsequent $\eta$ decays. The decay rate will become efficient much later than the chemical freeze-out because of the very small coupling $y \ll1$. The observed dark matter energy density is then given by 
\begin{equation}
    (\Omega_{\hbox{\tiny DM}}h^2)_{\hbox{\scriptsize obs.}}=  (\Omega_{\hbox{\tiny DM}}h^2)_{\hbox{\scriptsize freeze-in}} +  (\Omega_{\hbox{\tiny DM}}h^2)_{\hbox{\scriptsize super-WIMP}} \, ,
    \label{energy_density_tot}
\end{equation}
where $ (\Omega_{\hbox{\tiny DM}}h^2)_{\hbox{\scriptsize obs.}}=0.1200 \pm 0.0012$~\cite{Aghanim:2018eyx}. 

\begin{figure}[t!]
    \centering
    \includegraphics[scale=0.55]{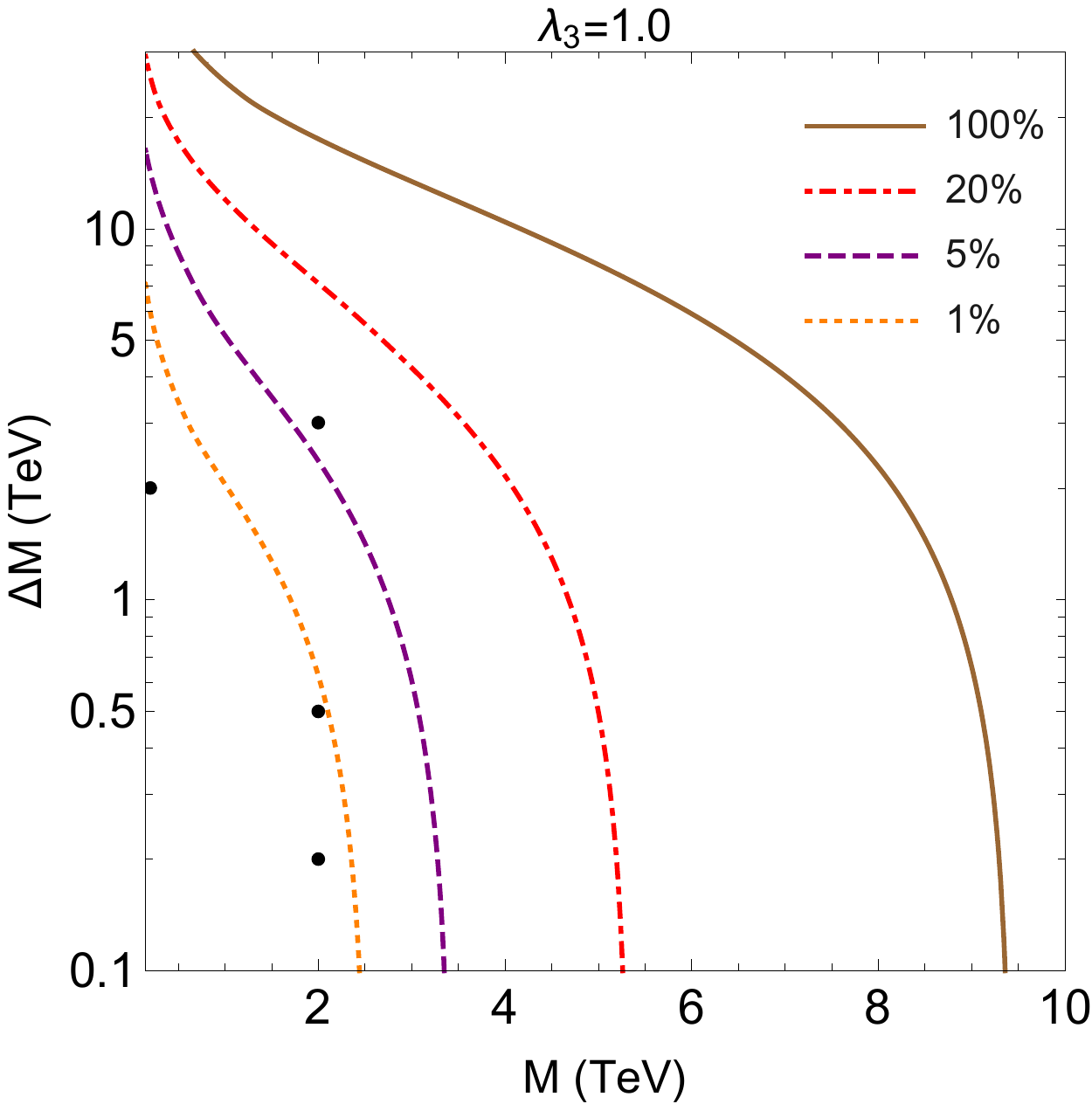}
    \hspace{0.9 cm}
    \includegraphics[scale=0.63]{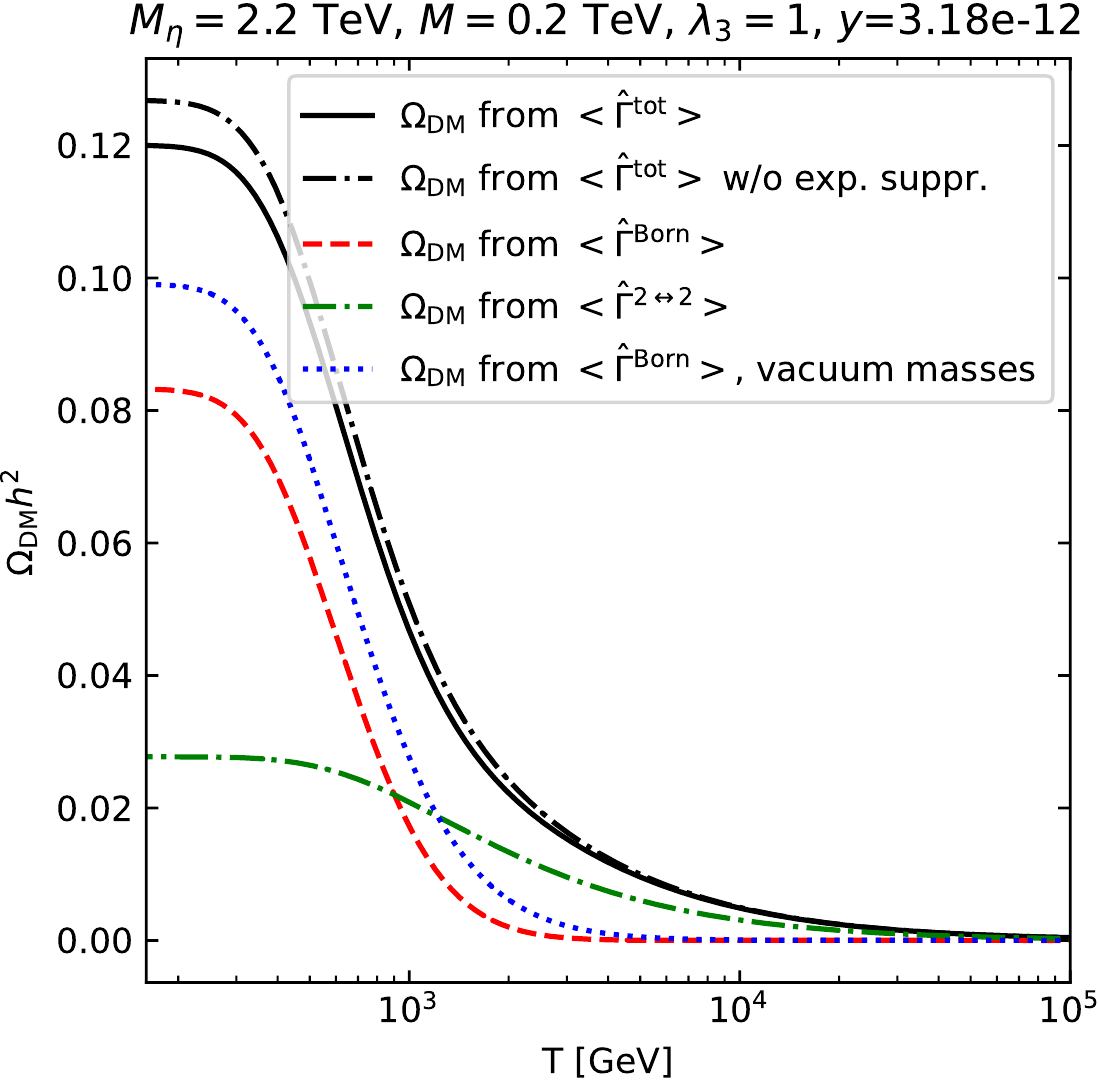}
    \caption{Left panel: the curves reproduce various fractions of the dark matter energy density upon using the thermally averaged cross section comprising Sommerfeld enhancement and bound state effects, for $\lambda_3=1.0$. We show four benchmark points, black dots, for which we discuss the freeze-in production. Right-panel: Dark matter energy density for the benchmark point P1 with $\lambda_3=1.0$. Different contributions corresponding to different rates are shown.}
    \label{fig:super_WIMP}
\end{figure}

Despite our focus is on the freeze-in production, a systematic assessment of the super-WIMP contribution is necessary. Indeed, it is important to understand which of the two production process is dominant in the parameter space of the model. The dynamics of the freeze-out and later-stage annihilations of colored scalars, as part of a dark matter model, has been extensively studied, see e.g.~\cite{Edsjo:1997bg,deSimone:2014pda,Ellis:2014ipa,Liew:2016hqo,Kim:2016kxt,Biondini:2018pwp,Biondini:2018ovz,Harz:2018csl}. The main outcome is that QCD gluon exchanges induce a Sommerfeld enhancement along with bound-state formation for the non-relativistic colored pairs (in the case of attractive channels), thus substantially reducing the abundance of the $\eta$'s with respect to the free annihilation cross section. For our discussion, we shall adapt the treatment of the present model as presented in ref.~\cite{Biondini:2018pwp} by taking the limit of very small $y$'s. In this approach non-relativistic effective field theories are exploited. One determines a thermally modified QCD potential and solve the corresponding Schr\"odinger equation, that provides thermally averaged Sommerfeld factors in terms of the spectral function of the colored scalar pair. It is worth remarking that, with this approach, the annihilating pair is determined dynamically, and bound states appear naturally when decreasing the temperature of the plasma (this is pretty much the same of the sequential melting seen in hevay-ion collisions for heavy-quarkonium systems). With respect to the analyses carried out in ref.~\cite{Garny:2018ali}, we include bound-state effects in our work, that further boost the scalar annihilations in addition to the above-threshold Sommerfeld enhancement. The result is shown in figure~\ref{fig:super_WIMP} (left panel), where the curves corresponding to different fractions of the super-WIMP contribution to the overall observed energy density are displayed in the plane $(M, \Delta M)$. The black dots are points in the parameter space where the freeze-in contribution is largely dominant.

In the following, we show the final results on the dark matter energy density when using different production rates discussed in the former sections. We present here the results for three benchmark points (for which the super-WIMP contribution to $ (\Omega_{\hbox{\tiny DM}}h^2)_{\textrm{obs.}}$ is less than 1\%)
\\
\begin{center} P1 ($M=0.2$ TeV, $M_\eta=2.2$), P2 ($M=2.0$ TeV, $M_\eta=2.5$) and P3 ($M=2.0$ TeV, $M_\eta=2.2$),
\end{center}
in the case of a top-like quark, with running Yukawa coupling $h_q$.
The rate equation reads, again from (\ref{diff_rate_eq}) upon defining the yield $Y=n_\mathrm{DM}/s$, as follows 
\begin{equation}
\label{yevolution}
    \frac{d\, Y}{d\,x}=2\frac{\langle \hat\Gamma\rangle}{s},
\end{equation}
where $x\equiv\ln(T_\mathrm{max}/T)$, with $T_\mathrm{max}$ the temperature
where we start the evolution, with $Y(x=0)=0$. 
$\hat O\equiv O/(3c_s^2 H)$, with $c_s^2$
the speed of sound squared and $H$ is the Hubble rate. Finally, $\langle\ldots\rangle\equiv \int_{\bf k} \ldots n_\mathrm{F}(k^0)$.
In what follows, we will use the parametrizations of \cite{Laine:2015kra}
for the speed of sound and entropy and energy densities, the latter entering
the Hubble rate.

\begin{figure}[t!]
    \centering
    \includegraphics[width=7.4 cm,height=7.4 cm]{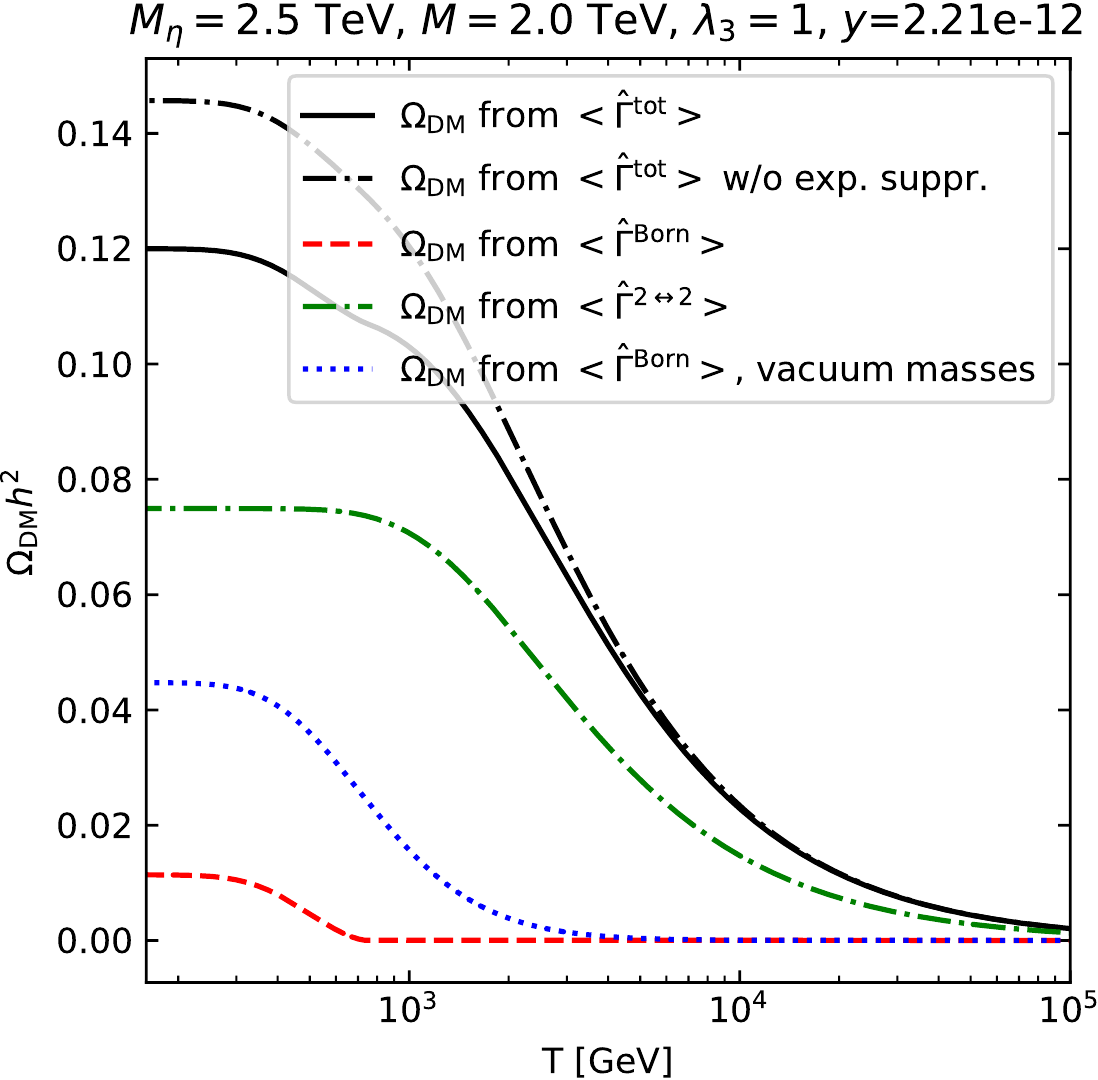}
    \hspace{0.9 cm}
    \includegraphics[width=7.4 cm,height=7.4 cm]{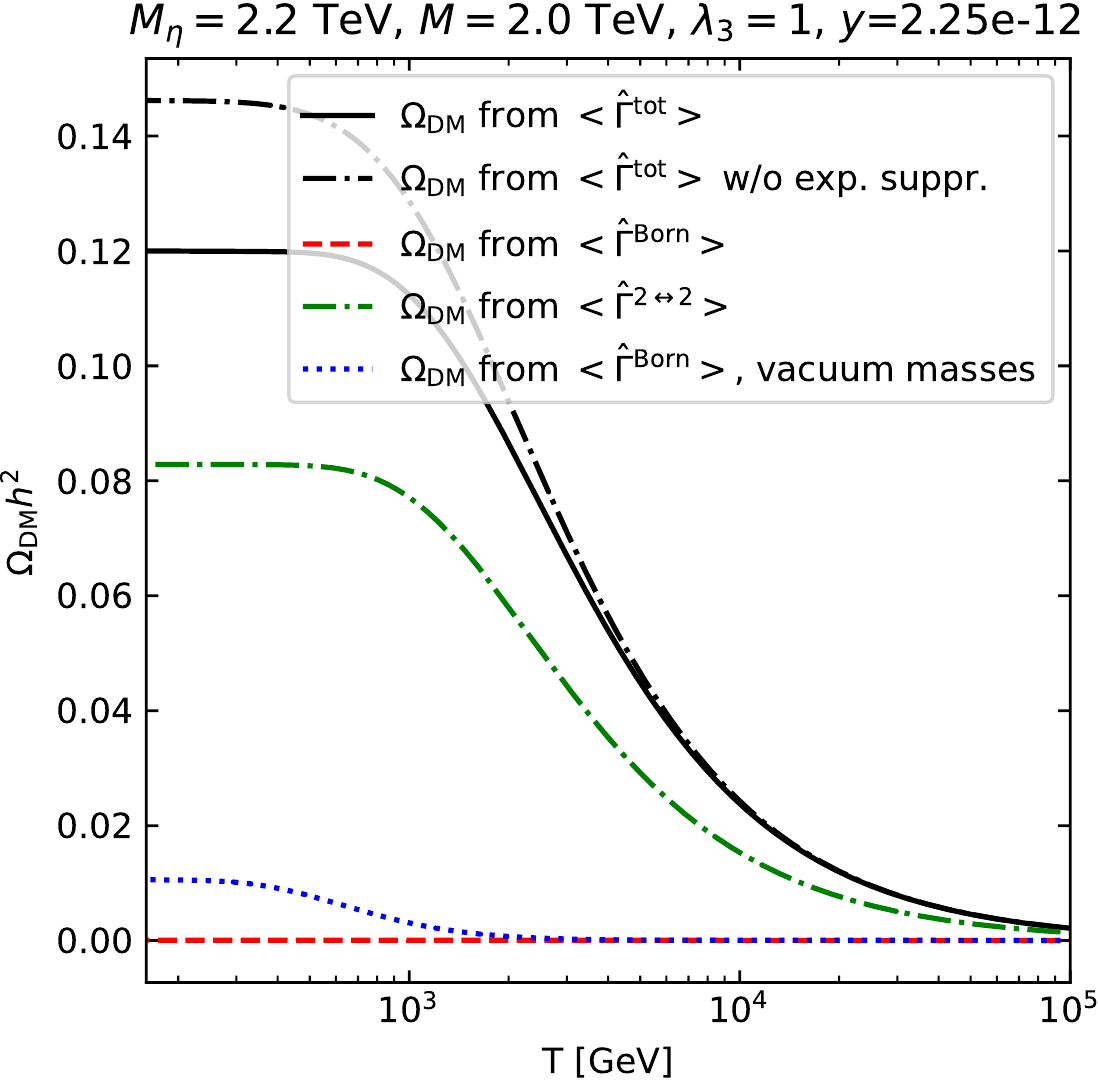}
    \caption{Dark matter energy density for the benchmark point P2 and P3 with $\lambda_3=1.0$. The up-quark Yukawa is $h_t$, running and non-vanishing.}
    \label{fig_energy_density_P3_comp}
\end{figure}

We do not perform a systematic scan of the parameter space. However, as a general trend, we find that the smaller the relative mass splitting $\Delta M /M$, the larger the impact of thermal masses, LPM and $2 \leftrightarrow 2$ processes. This can be understood in terms of the decreasing phase space in the $\eta \to \chi q$ decay process as $\Delta M / M$ gets smaller. The effect of neglecting the high-temperature contribution to the DM prodution is rather striking, as one can see in figure~\ref{fig:super_WIMP} (right panel) and \ref{fig_energy_density_P3_comp}. Even in the case of the largest mass splitting considered here, $\Delta M / M =10$,  high-temperature contributions still gives a 20\% (40\%) correction to the in-vacuum (thermal-mass included) Born production. For the smaller mass splitting in our analysis, $\Delta M / M =0.1$ the correction are well beyond $\mathcal{O}(1)$.  

\section*{4. Conclusions}
\label{sec:concl}
In this conference paper we summarize the finding of the work \cite{Biondini:2020ric}. The main scope of the study is the assessment of the high-temperature contributions to the freeze-in production mechanism, as induced by renormalizable operators (see Lagrangian (\ref{Lag_RT})). For this scenario, the bulk of the dark matter production is expected to happen for $T \sim M$, so to define what is usually dubbed in the literature as infra-red freeze-in. However, the results shown here, and discussed in more detail in ref.\cite{Biondini:2020ric}, demonstrate that this is not always the case: the dark matter population, as produced in the ultrarelativistic regime, can largely contribute to the total observed energy density. This is due to very efficient processes that take place in a high-temperature plasma, where all particles can be regarded as effectively massless and a collinear kinematic is established. We addressed multiple soft scatterings, that enhance the $1 \to 2$ decay process and make others viable, as well as $2 \to 2$ scatterings. For this model, the contributions to the DM energy density from such high-temperature processes induce corrections to the in-vaccum rate that range from $\mathcal{O}(20\%)$ up to from $\mathcal{O}(10)$, depending on the mass splitting $\Delta M$. The main theoretical improvement that can be foreseen is the derivation of NLO rates for $T \sim M_\eta,M$. This calculation would indeed allows us to ameliorate and make smoother the matching of the rates from the ultra-relativistic to the low-temperature regime. 

As for models that feature interactions between the accompanying state in the dark sector and the Standard Model (see e.g. refs. \cite{An:2013xka,Biondini:2019int,Belanger:2018sti,Arina:2020udz,Hisano:2011cs,DiFranzo:2013vra,Garny:2015wea,Junius:2019dci,Belanger:2018ccd}), our findings suggest a careful reassessment of the cosmologically preferred parameter space. The observed energy density is indeed a cornerstone of dark matter phenomenology, and one needs to map out the parameter space of the model that reproduces  $(\Omega_{\hbox{\tiny DM}}h^2)_{\hbox{ \scriptsize obs.}}$, which is by the way precises at 1\% level. 

\section*{Acknowledgements}
I am grateful to the Organizers of the BSM 2021 conference for the opportunity to present the research work, and to Jacopo Ghiglieri for the collaboration on the work presented at the conference. 

\bibliographystyle{unsrt}
\bibliography{Biondini_S.bib}
\end{document}